\documentclass[11pt,a4paper,epsfig]{article}
\usepackage{amsfonts}
\usepackage{epsfig}

\title{{\bf One-loop mass shift formula}\\{\bf for} \\ {\bf kinks and self-dual vortices}}

\author{A. Alonso Izquierdo$^{(a)}$,
W. Garcia Fuertes$^{(b)}$\\ M. de la Torre Mayado$^{(c)}$, J.
Mateos Guilarte$^{(d)}$
\\ {\normalsize {\it $^{(a)}$ Departamento de Matematica
Aplicada}, {\it Universidad de Salamanca, SPAIN}}\\{\normalsize
{\it $^{(b)}$ Departamento de Fisica} ,{\it Universidad de Oviedo,
SPAIN}}\\ {\normalsize {\it $^{(c)}$ Departamento de Fisica} ,{\it
Universidad de Salamanca, SPAIN}}\\{\normalsize{\it $^{(d)}$
Departamento de Fisica and IUFFyM}, {\it Universidad de Salamanca,
SPAIN}}}
\date{}

\begin{document}
\maketitle
\begin{abstract}
A formula is derived that allows us to compute one-loop mass
shifts for kinks and self-dual Abrikosov-Nielsen-Olesen vortices.
The procedure is based in canonical quantization and heat
kernel/zeta function regularization methods.
\end{abstract}

\section{{\bf Introduction}}

Abrikosov vortex lines \cite{Abri} were rediscovered by Nielsen
and Olesen in the realm of the Abelian Higgs model and were
proposed as models for dual strings \cite{NO}. In this framework,
the interest of studying the quantum nature of these
quasi-one-dimensional extended structures was immediately
recognized; contrarily to the macroscopic Ginzburg-Landau theory
of Type II superconductors, the birth-place of magnetic flux
tubes, the Abelian Higgs model is expected to play a r$\hat{{\rm
o}}$le in microscopic physics. This issue was first addressed in
Section \S. 3 of the original Nielsen-Olesen paper; taking the
zero thickness limit of the vortex line, the quantization
techniques of the old string theory were applied.

In this communication we shall deal with one-loop mass shifts for
the topological solitons that generate the thick string
structures. The mass of the topological solitons of the
(2+1)-dimensional Abelian Higgs model become the string tension of
the flux tubes embedded in three dimensions. Thus, from the
(3+1)-dimensional perspective, semi-classical string tensions will
be considered. In particular we offer as a novelty the derivation
of the vortex Casimir energy from the canonical quantization of
the planar Abelian Higgs model. With this demonstration, we shall
arrive at the starting point chosen in References \cite{AJmW1} and
\cite{AJmW2} to derive a formula for the one-loop vortex mass
shifts. The formula involves the coefficients of the heat-kernel
expansion associated with the second-order fluctuation operator
and affords us a numerical computation of the mass shifts. Before
our work, only fermionic fluctuations on vortex backgrounds have
been accounted for by Bordag in \cite{Bor}.

The control of the ultra-violet divergences arising in the
procedure will be achieved by using heat kernel/zeta function
regularization methods. In the absence of detailed knowledge of
the spectrum of the differential operator governing second-order
fluctuations around vortices, the expansion of the associated heat
kernel will be used in a way akin to that developed in the
computation of one-loop mass shifts for one-dimensional kinks, see
\cite{AJMW1}, \cite{AJMW2}, \cite{AJMW3}. In fact, a similar
technique has been applied before to compute the mass shift for
the supersymmetric kink \cite{BRvNV}, although in this latter case
the boundary conditions must respect supersymmetry. In the case of
vortices, the only available results besides the work reported
here refer to supersymmetric vortices and were achieved by
Vassilevich and the Stony Brook/Wien group, \cite{Vas},
\cite{RvNW}.

\section{High-temperature one-loop kink mass shift formula}
We start by very briefly treating the parallel and simpler
development for the kink of the $\lambda(\phi)^4_2$-model. With
the conventions of \cite{AJMW1} the one-loop kink mass shift
formula in the $\lambda(\phi)^4_2$ model on a line is formally: $
\Delta M_K=\Delta M_K^C+\Delta M_K^R$. The two pieces are:
\begin{enumerate}

\item The kink Casimir energy measured with respect to the vacuum
Casimir energy -zero point energy renormalization-:
\begin{equation}
\Delta M_K^C=\Delta E(\phi_K)- \Delta E_0=\frac{\hbar
m}{2\sqrt{2}}\left({\rm Tr}K^{{1\over 2}}-{\rm Tr}K_0^{{1\over
2}}\right) \label{eq:cas}
\end{equation}
\[
K=-{d^2\over dx^2}+4-{6\over{\rm cosh}^2x} \hspace{2cm} ,
\hspace{2cm} K_0=-{d^2\over dx^2}+4 \qquad .
\]
$K$ and $K_0$ are the differential operators governing the
second-order fluctuations around the kink and the vacuum
respectively.

\item The contribution of the mass renormalization counter-terms to the one-loop kink
mass:
\begin{equation}
\Delta M_K^R=-3\frac{\hbar m}{\sqrt{2}}\cdot I(4)\cdot \int \, dx
\,\left(\phi_K^2(x)-\phi_\pm^2\right) \hspace{0.5cm},
\hspace{0.5cm} I(4)=\int \, \frac{d^2k}{(2\pi)^2} \, \cdot
\frac{i}{(k_0^2-k^2-4+i\varepsilon)} \qquad .\label{eq:ren}
\end{equation}
\end{enumerate}
The ultraviolet divergences are regularized by means of
generalized zeta functions:
\[
\Delta M_K^C(s)={ \hbar\over 2} (2{\mu^2\over m^2})^s\mu
\left(\zeta_K(s)-\zeta_{K_0}(s)\right)\quad , \quad \Delta
M_K^R(s)=-\frac{6\hbar}{L}\cdot\left(\frac{2\mu^2}{m^2}\right)^{s+{1\over
2}}\cdot\frac{\Gamma(s+1)}{\Gamma(s)}\cdot\zeta_{K_0}(s+1) \quad ,
\]
where $s$ is a complex parameter; $\mu$ a parameter of dimension
$L^{-1}$, and $L$ is a normalization length on the line. From the
partition functions, one obtains the generalized zeta functions
via Mellin transform:
\[
{\rm Tr}e^{-\beta K_0}=\frac{mL}{\sqrt{8\pi\beta}}\cdot
e^{-4\beta} \hspace{0.5cm} , \hspace{0.5cm} {\rm Tr}^* e^{-\beta
K}=\frac{mL}{\sqrt{8\pi\beta}}\cdot e^{-4\beta}+e^{-3\beta}(1-{\rm
Erfc}\sqrt{\beta})-{\rm Erfc}2\sqrt{\beta}
\]
\[
\zeta_{K_0}(s)={mL\over\sqrt{8\pi}}\cdot \frac{\Gamma(s-{1\over
2})}{2^{2s-1}\Gamma(s)} \qquad ; \qquad
\zeta^*_{K}(s)=\zeta_{K_0}(s)+\frac{\Gamma(s+{1\over
2})}{\sqrt{\pi}\Gamma(s)}\left[{2\over 3^{s+{1\over 2}}}\cdot {}_2
F_1[{1\over 2},s+{1\over 2},{3\over 2},-{1\over 3}]-{1\over
4^s}{1\over s}\right] \quad ,
\]
passing from complementary error functions ${\rm Erfc}x$ to
hypergeometric Gauss functions ${}_2F_1[a,b,c;d]$. The star means
that the zero mode is not accounted for. Because $\Delta
M_K^C=\lim_{s\rightarrow -{1\over 2}}\Delta M_K^C(s)$ and $\Delta
M_K^R=\lim_{s\rightarrow -{1\over 2}}\Delta M_K^R(s)$
\[
\Delta M_K^C= -\frac{\hbar
m}{2\sqrt{2}\pi}\lim_{\varepsilon\rightarrow
0}\left[{3\over\varepsilon}+3\ln\frac{2\mu^2}{m^2}-\frac{\pi}{\sqrt{3}}\right]
\quad , \quad \Delta M_K^R=\frac{\hbar
m}{2\sqrt{2}\pi}\lim_{\varepsilon\rightarrow
0}\left[{3\over\varepsilon}+3\ln\frac{2\mu^2}{m^2}-2(2+1)\right]
\]
provides the exact Dashen-Hasslacher-Neveu (DHN) result, see
\cite{AJMW1} and References quoted therein:
\begin{center}
{\normalsize \begin{tabular}{|c|} \hline \\   $\Delta M_K=\Delta
M_K^C+\Delta M_K^R=\frac{\hbar m}{2\sqrt{6}}-\frac{3\hbar
m}{\pi\sqrt{2}}$
\\ \\
\hline
\end{tabular}}\qquad .
\end{center}
Without using the knowledge of the spectrum of $K$, one can rely
on the high-temperature expansion of the partition function:
\[
{\rm Tr}e^{-\beta K}=\frac{e^{-4\beta}}{\sqrt{4\pi\beta}}\cdot
\sum_{n=0}^\infty \, c_n(K)\, \beta^n  \quad , \quad
c_0(K)=\lim_{L\rightarrow\infty}\frac{mL}{\sqrt{2}}\hspace{0.5cm},\hspace{0.5cm}
c_{n}(K)=\frac{2^{n+1}(1+2^{2n-1})}{(2 n-1)!!}\, , \hspace{0.5cm}
n\geq 1
\]
to find:
\[
\zeta_K(s)= \frac{1}{\Gamma(s)}\left[ {1\over\sqrt{4\pi}}\cdot
\sum_{n=0}^{\infty}c_n(K)\cdot \frac{\gamma[s+n-{1\over
2},4]}{4^{s+n-{1\over 2}}}-\int_0^1 \, d\beta \,
\beta^{s-1}\right]+\int_1^\infty \, d\beta \, \beta^{s-1} {\rm
Tr}^* e^{-\beta K} \qquad .
\]
Here, the zero mode has been subtracted and the meromorphic
structure of $\zeta_K(s)$ is encoded in the incomplete Gamma
functions $\gamma[s+n-{1\over 2},4]$. Neglecting the (very small)
contribution of the entire function, and cutting the series at a
large but finite $N_0$, the kink Casimir energy becomes:
\[
\Delta M_K^C \simeq \frac{\hbar}{2}\cdot \lim_{s\rightarrow
-{1\over 2}}\left(\frac{2\mu^2}{m^2}\right)^s\cdot \mu \cdot
\frac{1}{\Gamma(s)}\cdot
\left[{1\over\sqrt{4\pi}}\sum_{n=1}^{N_0}c_n(K)\frac{\gamma[s+n-{1\over
2},4]}{4^{s+n-{1\over 2}}} -{1\over s}\right] \qquad ,
\]
i.e. the zero-point vacuum energy renormalization takes care of
the term coming from $c_0(K)$. Note that
$\zeta_{K_0}(s)\simeq\frac{mL}{\sqrt{8\pi}}\cdot\frac{\gamma[s-{1\over
2},4]}{2^{2s-1}\Gamma(s)}$ in the $\beta<1$ regime where the
high-$T$ expansion is reliable. The other correction due to the
mass renormalization counter-terms can be arranged also into
meromorphic and entire parts:
\[
\Delta M_K^R=-\frac{\hbar \mu}{2\sqrt{4\pi}}\cdot c_1(K) \cdot
\lim_{s\rightarrow -{1\over
2}}\left(\frac{2\mu^2}{m^2}\right)^{s+{1\over 2}}\cdot
\frac{1}{4^{s+{1\over 2}}\Gamma(s)}\cdot  \left[\gamma[s+{1\over
2},4]+\Gamma[s+{1\over 2},4] \right]
\]
The mass renormalization term exactly cancels the $c_1(K)$
contribution. Our minimal subtraction scheme fits with the
renormalization prescription set in \cite{Beklv}: for theories
with only massive fluctuations, the quantum corrections should
vanish in the limit in which all masses go to infinity. This
criterion requires precisely the cancelation found. We end with
the high-temperature one-loop kink mass shift formula:
\begin{center}
{\normalsize \begin{tabular}{|c|} \hline \\   $\Delta
M_K=-\frac{\hbar m}{4\sqrt{2\pi}}\cdot
\left[\frac{1}{\sqrt{4\pi}}\cdot\sum_{n=2}^{N_0}\, c_n(K)\cdot
\frac{\gamma[n-1,4]}{4^{n-1}}+2\right]\quad$,$\qquad
\beta=\frac{\hbar m}{k_B T}<1$
\\ \\
\hline
\end{tabular}}
\end{center}
Finally, by applying this formula with $N_0=11$ we have:
\[
\Delta M_K \cong -0.471371 \hbar m \qquad ,
\]
with an error with respect to the DHN result of: $0.0002580 \hbar
m$.
\section{{\bf The planar Abelian Higgs model}}
In this Section we generalize formulae (\ref{eq:cas}) and
(\ref{eq:ren}) to determine the one-loop mass shift of vortices in
the Abelian Higgs model. We shall derive the formula that serves
as the starting point in papers \cite{AJmW1} and \cite{AJmW2},
thus filling a gap in the issue. Within the conventions stated in
these References, we write the action governing the dynamics of
the AHM in the form:
\[
S= \frac{v}{e}\int d^3 x \left[ -\frac{1}{4} F_{\mu \nu} F^{\mu
\nu}+\frac{1}{2} (D_\mu \phi)^* D^\mu \phi - \frac{\kappa^2}{8}
(\phi^* \phi-1)^2 \right] \qquad .
\]
A shift of the complex scalar field from the vacuum
$\phi(x^\mu)=1+H(x^\mu)+iG(x^\mu)$ and choice of the Feynman-'t
Hooft renormalizable gauge $R(A_\mu , G)=\partial_\mu A^\mu(x^\mu)
- G(x^\mu)$ lead us to write the action in terms of Higgs $H$,
Goldstone $G$, vector boson $A_\mu$ and ghost $\chi$ fields:
\begin{eqnarray*}
S+S_{{\rm g.f.}}+S_{{\rm ghost}}&=&{v\over e}\int \, d^3x \,
\left[ -\frac{1}{2} A_\mu
[-g^{\mu\nu}(\Box +1)]A_\nu +\partial_\mu\chi^*\partial^\mu \chi- \chi^*\chi
+\frac{1}{2}\partial_\mu G\partial^\mu G-\frac{1}{2} G^2 \right.\\
&+&\frac{1}{2}\partial_\mu H\partial^\mu H-\frac{\kappa^2}{2}
H^2-{\kappa^2\over 2}H (H^2+G^2)+H (A_\mu A^\mu -\chi^*
\chi)\\&+&\left. A_\mu (\partial^\mu H G-\partial^\mu G
H)-\frac{\kappa^2}{8} (H^2+G^2)^2 +\frac{1}{2}(G^2+H^2) A_\mu
A^\mu \right] \qquad .
\end{eqnarray*}

\subsection{{\bf Vacuum Casimir energy}}
Canonical quantization promoting the coefficients of the plane
wave expansion around the vacuum of the fields to operators
provides the free quantum Hamiltonian:
\begin{itemize}
\item If $m=ev$,
\[
\delta A_\mu(x_0,\vec{x})=\left(\frac{\hbar^{{1\over
2}}}{e^{{1\over 2}}v^{{3\over 2}}L}\right) \cdot
\sum_{\vec{k}}\sum_{\alpha}\frac{1}{\sqrt{2\omega(\vec{k})}}\left[
a^*_\alpha(\vec{k})e^\alpha_\mu(k)e^{ikx}+a_\alpha(\vec{k})e^\alpha_
\mu(k)e^{-ikx}\right]
\]
\[
[\hat{a}_\alpha(\vec{k}),\hat{a}_\alpha^\dagger
(\vec{q})]=(-1)^{\delta_{\alpha
0}}\delta_{\alpha\beta}\delta_{\vec{k}\vec{q}} \Rightarrow
H^{(2)}[\delta \hat{A}_\mu]=\sum_{\vec{k}}\sum_\alpha \hbar m
\omega(\vec{k})\left((-1)^{\delta_{\alpha
0}}\hat{a}_\alpha^\dagger(\vec{k})\hat{a}_\alpha(\vec{k})+{1\over
2})\right) \quad .
\]
\item
\[
\delta
H(x_0,\vec{x})=\frac{1}{vL}\sqrt{\frac{\hbar}{ev}}\sum_{\vec{k}}\frac{1}{\sqrt{2\nu(\vec{k})}}\left[a^*(\vec{k})e^{ikx}+a(\vec{k})e^{-ikx}\right]
\quad , \quad \nu(\vec{k})=+\sqrt{\vec{k}\vec{k}+\kappa^2}
\]
\[
[\hat{a}(\vec{k}),\hat{a}^\dagger(\vec{q})]=\delta_{\vec{k}\vec{q}}\Rightarrow
H^{(2)}[\delta\hat{H}]=\hbar
m\sum_{\vec{k}}\omega(k)\left(\hat{a}^\dagger(\vec{k})\hat{a}(\vec{k})+{1\over
2}\right) \quad .
\]
\item
\[
\delta
G(x_0,\vec{x})=\frac{1}{vL}\sqrt{\frac{\hbar}{ev}}\sum_{\vec{k}}\frac{1}{\sqrt{2\omega(\vec{k})}}\left[b^*(\vec{k})e^{ikx}+b(\vec{k})e^{-ikx}\right]
\quad , \quad \omega(\vec{k})=+\sqrt{\vec{k}\vec{k}+1}
\]
\[
[\hat{b}(\vec{k}),\hat{b}^\dagger(\vec{q})]=\delta_{\vec{k}\vec{q}}\Rightarrow
H^{(2)}[\delta\hat{H}]=\hbar
m\sum_{\vec{k}}\omega(k)\left(\hat{b}^\dagger(\vec{k})\hat{b}(\vec{k})+{1\over
2}\right) \quad .
\]
\item Canonical quantization proceeds by anti-commutators for
ghost  fields
\[
\delta
\chi(x_0,\vec{x})=\frac{1}{vL}\sqrt{\frac{\hbar}{ev}}\sum_{\vec{k}}\frac{1}{\sqrt{2\omega(\vec{k})}}\left[c(\vec{k})e^{ikx}+d^*(\vec{k})e^{-ikx}\right]
\]
\[
\{\hat{c}^\dagger(\vec{k}),\hat{c}(\vec{q})\}=\{\hat{d}^\dagger(\vec{k}),\hat{d}(\vec{q})\}=\delta_{\vec{k}\vec{q}}
\Rightarrow H^{(2)}[\delta\hat{\chi}]=\hbar
m\sum_{\vec{k}}\omega(k)\left(\hat{c}^\dagger(\vec{k})\hat{c}(\vec{k})+\hat{d}^\dagger(\vec{k})\hat{d}(\vec{k})-1\right)\quad
.
\]
\end{itemize}
The vacuum Casimir energy is the sum of four contributions: if
$\bigtriangleup=\sum_{j=1}^2 \, \frac{\partial}{\partial x_j}\cdot
\frac{\partial}{\partial x_j}$ denotes the Laplacian,
\[
\Delta E^{(1)}_0=\sum_{\vec{k}}\sum_\alpha {\hbar m\over 2}
\omega(\vec{k})=\frac{3\hbar m}{2}{\rm
Tr}[-\bigtriangleup+1]^{{1\over 2}} \, \, , \, \, \Delta
E^{(2)}_0=\sum_{\vec{k}}{\hbar m\over 2} \nu(\vec{k})=\frac{\hbar
m}{2}{\rm Tr}[-\bigtriangleup+\kappa^2]^{{1\over 2}}
\]
\[
\Delta E^{(3)}_0=\sum_{\vec{k}}{\hbar m\over 2}
\omega(\vec{k})=\frac{\hbar m}{2}{\rm
Tr}[-\bigtriangleup+1]^{{1\over 2}}\quad , \quad \Delta
E^{(4)}_0=-\sum_{\vec{k}}\hbar m \omega(\vec{k})=-\hbar m{\rm
Tr}[-\bigtriangleup+1]^{{1\over 2}}
\]
come from the vacuum fluctuations of the vector boson, Higgs,
Goldstone and ghost fields. Ghost fluctuations, however, cancel
the contribution of temporal vector bosons and Golstone particles,
and the vacuum Casimir energy in the planar AHM is due only to
Higgs particles and transverse massive vector bosons:
\[
\Delta E_0=\sum_{r=1}^4 \Delta E^{(r)}_0=\hbar m{\rm
Tr}[-\bigtriangleup+1]^{{1\over 2}}+{\hbar m\over 2}{\rm
Tr}[-\bigtriangleup+\kappa^2]^{{1\over 2}} \qquad .
\]

\subsection{{\bf Vortex Casimir energy}}
At the critical point between Type I and Type II
superconductivity, $\kappa^2=1$, the energy can be arranged in a
Bogomolny splitting:
\[
E=v^2 \int \frac{d^2 x}{2} \left( |D_1 \phi \pm i D_2 \phi|^2 + [
F_{12} \pm {\textstyle\frac{1}{2}} (\phi^* \phi-1) ]^2
\right)+\frac{1}{2}v^2|g| \quad , \quad g= \int d^2 x F_{12}=2{\pi
l} \, \, , \, \, l\in{\mathbb Z} \quad .
\]
Therefore, the solutions of the first-order equations
\[
D_1 \phi \pm i D_2 \phi=0 \hspace{0.3cm};\hspace{0.3cm} F_{12} \pm
\frac{1}{2} (\phi^*\phi-1) =0
\]
are absolute minima of the energy, hence stable, in each
topological sector with a classical mass proportional to the
magnetic flux. Assuming a purely vorticial vector field plus the
spherically symmetric ansatz
\begin{eqnarray*}
\phi_1(x_1,x_2) = f(r) {\rm cos}l\theta \quad &,& \quad
\phi_2(x_1,x_2) = f(r) {\rm sin}l\theta \\
A_1(x_1,x_2) =-l \frac{\alpha(r)}{r}{\rm sin}\theta \quad &,&
\quad A_2(x_1,x_2) = l \frac{\alpha(r)}{r}{\rm cos}\theta \quad ,
\end{eqnarray*}
$g= - \oint_{r=\infty} dx_i A_i = -l\oint_{r=\infty}{
[x_2dx_1-x_1dx_2]\over r^2}=2 \pi l$, the first-order equations
reduce to
\[
{1\over r} {d \alpha \over d r}(r)= \mp \frac{1}{2 l} (f^2(r)-1)
\qquad , \qquad {d f\over d r}(r) = \pm \frac{l}{r}
f(r)[1-\alpha(r)] \qquad ,
\]
to be solved together with the boundary conditions $
{\displaystyle \lim_{r\rightarrow\infty}} f(r) = 1 \, , \,
{\displaystyle \lim_{r\rightarrow\infty}}\alpha(r) = 1 \, , \,
f(0) =0 \, , \, \alpha(0)=0 $ required by energy finiteness plus
regularity at the origin (center of the vortex). A partly
numerical, partly analytical procedure provides the field profiles
$f(r)$, $\alpha(r)$ as well as the magnetic field and the energy
density
\[
B(r)={l\over 2r}\frac{d\alpha}{dr} \hspace{1.5cm} , \hspace{1.5cm}
\varepsilon(r)={1\over 4}(1-f^2(r))^2+{l^2\over
r^2}(1-\alpha(r))^2 f^2(r) \qquad .
\]
plotted in Figure 1 for $l=1,2,3,4$.

\begin{center}
\includegraphics[height=2.0cm]{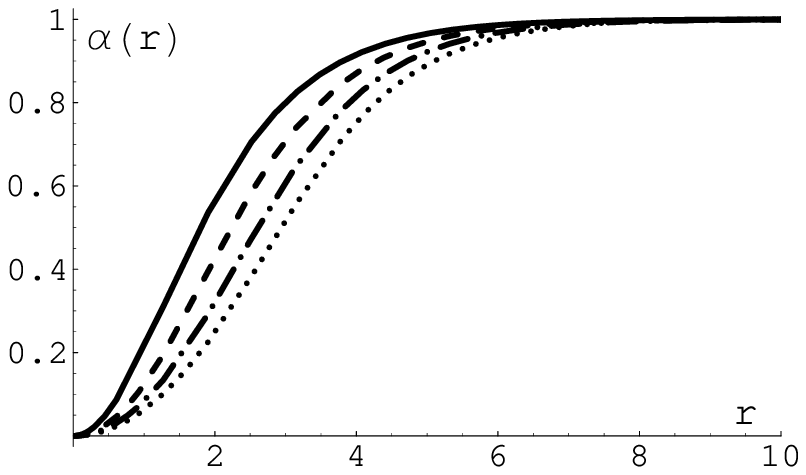}\hspace{0.8cm}
\includegraphics[height=2.0cm]{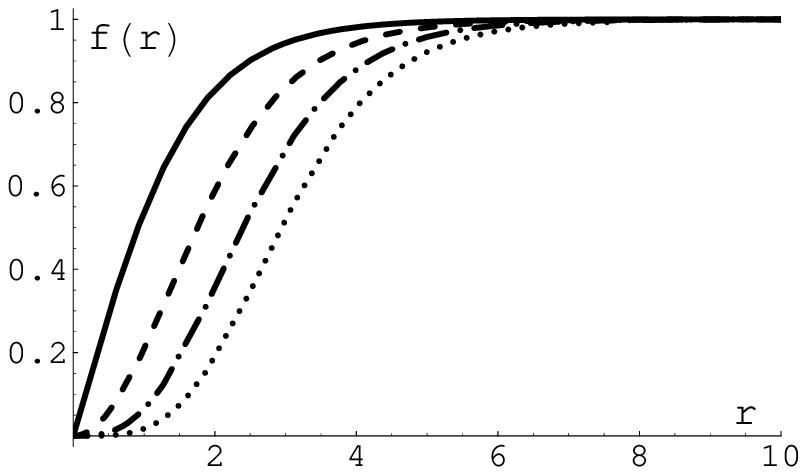}\hspace{0.8cm}
\includegraphics[height=2.0cm]{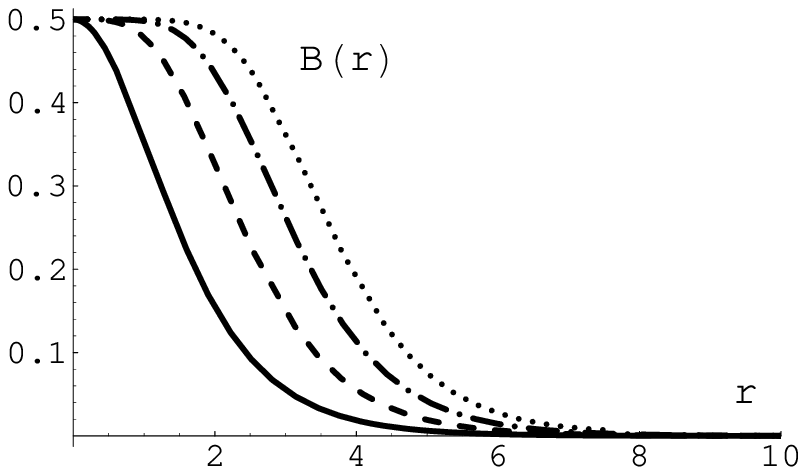}\hspace{0.8cm}
\includegraphics[height=2.0cm]{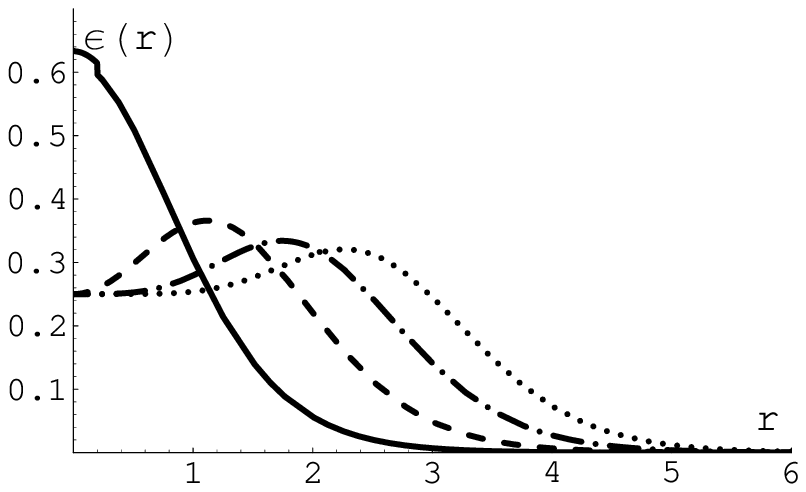} \\
{\small Figure 1. \textit{Plots of the field profiles $\alpha(r)$
(a) and $f(r)$ (b); the magnetic field $B(r)$ (c), and the energy
density $\varepsilon(r)$ for self-dual vortices with $l=1$ (solid
line), $l=2$ (dashed line), $l=3$ (broken-dashed line) and $l=4$
(dotted line).}}
\end{center}
Consider small fluctuations around vortices $ \phi(x_0,\vec{x})=
s(\vec{x})+\delta s(x_0,\vec{x}) \hspace{0.3cm} , \hspace{0.3cm}
A_k(x_0,\vec{x})=V_k(\vec{x})+\delta a_k (x_0,\vec{x})$, where by
$s(\vec{x})$ and $V_k(\vec{x})$ we respectively denote the scalar
and vector field of the vortex solutions. Working in the
Weyl/background gauge
\[
A_0(x_0,\vec{x})=0 \qquad \qquad , \qquad \qquad \partial_k\delta
a_k(x_0,\vec{x})+s_2(\vec{x})\delta
s_1(x_0,\vec{x})-s_1(\vec{x})\delta s_2(x_0,\vec{x})=0 \quad ,
\]
the classical energy up to ${\cal O}(\delta^2)$ order is:
\[
H^{(2)}+H^{(2)}_{{\rm g.f.}}+H^{(2)}_{{\rm ghost}}={v^2\over
2}\int \, d^2x\left\{\frac{\partial\delta\xi^T}{\partial
x_0}\frac{\partial\delta\xi}{\partial x_0}+\delta
\xi^T(x_0,\vec{x})K\delta\xi(x_0,\vec{x})+\delta\chi^*(\vec{x})\left(-\bigtriangleup
+|s(\vec{x})|^2\right)\delta\chi(\vec{x})\right\} \quad ,
\]
where
\[
\delta\xi(x_0,\vec{x})=\left(\begin{array}{c} \delta a_1 (x_0,\vec{x}) \\
\delta a_2 (x_0,\vec{x}) \\ \delta s_1(x_0,\vec{x}) \\ \delta
s_2(x_0,\vec{x})\end{array}\right) \qquad , \qquad
\nabla_js_a=\partial_js_a+\varepsilon_{ab}V_js_b \qquad ,
\]
and
\begin{eqnarray*}
K={\normalsize \left(\begin{array}{cccc} -\bigtriangleup +|s|^2 &
0 & -2\nabla_1s_2 & 2\nabla_1s_1
\\ 0 & -\bigtriangleup+|s|^2 & -2\nabla_2s_2 &
2\nabla_2s_1 \\ -2\nabla_1s_2 & -2\nabla_2s_2 &
-\bigtriangleup+{1\over 2}(3|s|^2+2V_kV_k-1) & -2V_k\partial_k
\\ 2\nabla_1s_1 & 2\nabla_2s_1 & 2V_k\partial_k  &
-\bigtriangleup+{1\over 2}(3|s|^2+2V_kV_k-1)
\end{array}\right)} \qquad .
\end{eqnarray*}

The general solutions of the linearized field equations
\[
\frac{\partial^2\delta\xi_A}{\partial
x_0^2}(x_0,\vec{x})+\sum_{B=1}^4 \,
K_{AB}\cdot\delta\xi_B(x_0,\vec{x})=0 \hspace{2cm} , \hspace{2cm}
K^G \delta\chi(\vec{x})=\left(-\bigtriangleup
+|s(\vec{x})|^2\right)\delta\chi(\vec{x})=0
\]
are the eigenfunction expansions
\[
\delta\xi_A^\prime (x_0,\vec{x})={1\over
vL}\sqrt{\frac{\hbar}{ev}}\cdot
\sum_{\vec{k}}\sum_{I=1}^4\frac{1}{\sqrt{2\varepsilon(\vec{k})}}\left[a^*_I(\vec{k})e^{i\varepsilon(\vec{k})
x_0}u^{(I)*}_A(\vec{x};\vec{k})+a_I(\vec{k})e^{-i\varepsilon(\vec{k})
x_0}u_A^{(I)}(\vec{x};\vec{k})\right]
\]
\[
\delta\chi^\prime (x_0,\vec{x})={1\over
vL}\sqrt{\frac{\hbar}{ev}}\cdot
\sum_{\vec{k}}\frac{1}{\sqrt{2\varepsilon^G(\vec{k})}}\left[c(\vec{k})u^*(\vec{x};\vec{k})+d^*(\vec{k})u(\vec{x};\vec{k})\right]
\qquad ,
\]
where $A=1,2,3,4$ and by $u^{(I)}(k)$, $u(k)$ the non-zero
eigenfunctions of $K$ and $K^G$ are denoted respectively:
$Ku^{(I)}(\vec{x})=\varepsilon(\vec{k})u^{(I)}(\vec{x})$,
$K^Gu(\vec{x})=\varepsilon^G(\vec{k})u(\vec{x})$. Canonical
quantization
\[
[\hat{a}_I(\vec{k}),\hat{a}_J^\dagger(\vec{q})]=\delta_{IJ}\delta_{\vec{k}\,\vec{q}}\quad
, \quad
\{\hat{c}(\vec{k}),\hat{c}^\dagger(\vec{q})\}=\delta_{\vec{k}\,\vec{q}}
\quad , \quad
\{\hat{d}(\vec{k}),\hat{d}^\dagger(\vec{q})\}=\delta_{\vec{k}\,\vec{q}}
\]
leads to the quantum free Hamiltonian
\[
\hat{H}^{(2)}+\hat{H}^{(2)}_{{\rm g.f.}}+\hat{H}^{(2)}_{{\rm
Ghost}}=\hbar m
\cdot\sum_{\vec{k}}\left[\sum_{I=1}^4\,\varepsilon(\vec{k})\left(\hat{a}_I^\dagger(\vec{k})\hat{a}_I(\vec{k})+{1\over
2}\right)+{1\over
2}\varepsilon^G(\vec{k})\left(\hat{c}^\dagger(\vec{k})\hat{c}(\vec{k})+\hat{d}^\dagger(\vec{k})\hat{d}(\vec{k})-1\right)\right]\quad
,
\]
and the vortex Casimir energy reads:
\[
\bigtriangleup E_V=\frac{\hbar m}{2}{\rm STr}^*\, K^{{1\over
2}}=\frac{\hbar m}{2}{\rm Tr}^*\, K^{{1\over 2}}-\frac{\hbar
m}{2}{\rm Tr}^*\, (K^{{\rm G}})^{{1\over 2}}  \qquad .
\]
Note that the ghost fields are static in this combined
Weyl-background gauge and their vacuum energy is one-half with
respect to the time-dependent case. Only the Goldstone
fluctuations around the vortices must be subtracted. The
zero-point vacuum energy renormalization provides an analogous
formula to (\ref{eq:cas}) for self-dual ($\kappa^2=1$) vortices
\begin{equation}
\bigtriangleup M_V^C=\bigtriangleup E_V-\bigtriangleup E_0={\hbar
m\over 2}\left[{\rm STr}^*\, K^{{1\over 2}}-{\rm STr}\,
K_0^{{1\over 2}}\right] \label{eq:casv}
\end{equation}

\subsection{{\bf Mass renormalization energy}}
Adding the counter-terms
\[
{\cal L}_{c.t.}^S = 2\hbar I(1) \left[|\phi|^2-1 \right] \quad ,
\quad {\cal L}_{c.t.}^A = -\hbar I(1) A_\mu A^\mu \hspace{0.5cm} ,
\hspace{0.5cm} I(1)=\int \, \frac{d^3k}{(2\pi)^3} \cdot
\frac{i}{k^2-1+i\varepsilon}
\]
to the Lagrangian, the divergences arising in the one-loop Higgs,
Goldstone and vector boson self-energy graphs as well as the Higgs
tadpole are exactly canceled. These terms add the following
contribution to the one-loop vortex mass shift:
\begin{equation}
\bigtriangleup M_V^R=\hbar m I(1) \int \, d^2x \,
\left[2(1-|s(\vec{x})|^2)-V_k(\vec{x})V_k(\vec{x})\right] \qquad ,
\label{eq:renv}
\end{equation}
and, formally, $\bigtriangleup M_V=\bigtriangleup
M_V^C+\bigtriangleup M_V^R$.

\section{{\bf High-temperature one-loop vortex mass shift
formula}}

As in the kink case, from the high-temperature expansion of the
heat kernels
\[
{\rm Tr}e^{-\beta K} =  {e^{-\beta}\over 4\pi\beta}\cdot
\sum_{n=0}^\infty \,\sum_{A=1}^4 \,\beta^n [c_n]_{AA}(K)
\hspace{1.5cm} , \hspace{1.5cm} {\rm Tr}e^{-\beta K^{\rm G}} =
{e^{-\beta}\over 4\pi\beta}\cdot \sum_{n=0}^\infty\beta^n
c_n(K^{\rm G})
\]
the vortex generalized zeta functions can be written in the form:
\[
\zeta_{K}(s)=\sum_{n=0}^\infty \sum_{A=1}^4
[c_n]_{AA}(K)\cdot\frac{\gamma[s+n-1,1]}{4\pi\Gamma(s)}+{1\over\Gamma(s)}\int_1^\infty
{\rm Tr}^* e^{-\beta K} \, d\beta
\]
\[
\zeta_{K^G}(s)=\sum_{n=0}^\infty
c_n(K^G)\cdot\frac{\gamma[s+n-1,1]}{4\pi\Gamma(s)}+{1\over\Gamma(s)}\int_1^\infty
\, d\beta \, {\rm Tr}^* e^{-\beta K^{G}} \qquad .
\]
Neglecting the entire part and setting a large but finite $N_0$
the vortex Casimir energy is regularized as
\[
\Delta M_V^C(s)=\frac{\hbar\mu}{2}\left({\mu^2\over
m^2}\right)^s\left\{-\frac{2l}{\Gamma(s)}\int_0^1 d\beta
\beta^{s-1}+\sum_{n=1}^{N_0}\left[\sum_{A=1}^4 \,
[c_n]_{AA}(K)-c_n(K^G)\right]\cdot
\frac{\gamma[s+n-1,1]}{\Gamma(s)}\right\} \qquad ,
\]
where the $2l$ zero modes have been subtracted: the zero-point
vacuum renormalization amounts to throwing away the contribution
of the $c_0(K)$ and $c_0(K^G)$ coefficients. Also, $\Delta M_V^R$
is regularized in a similar way
\[
\Delta M_V^R(s) = {\hbar\over 2\mu L^2}\left({\mu^2\over
m^2}\right)^s \zeta_{K^{\rm G}_0} (s) \Sigma
(s(\vec{x}),V_k(\vec{x})) \, ; \, \Sigma
(s(\vec{x}),V_k(\vec{x}))=\int  d^2x \, [2(1-|s(\vec{x})|^2)-
V_k(\vec{x})V_k(\vec{x})] \qquad .
\]
The physical limits $s=-{1\over 2}$ for $\Delta M_V^C$ and
$s={1\over 2}$ for $\Delta M_V^R$ are regular points of the zeta
functions (contrarily to the kink case). But, as in the kink case,
the contribution of the first coefficient of the asymptotic
expansion exactly kills the contribution of the mass
renormalization counter-terms. The explanation of this fact
proceeds along the same lines as in the kink case.
\[
\Delta M_V^{(1)C} (-1/2) = - {\hbar m \over 8 \pi } \Sigma (s,V_k)
\cdot {\gamma[-1/2,1] \over \Gamma(1/2)} \hspace{1cm} ,
\hspace{1cm} \Delta M_V^R (1/2) = {\hbar m \over 8 \pi} \cdot
\Sigma (s,V_k) \cdot
 {\gamma[-1/2,1]\over \Gamma(1/2)}
\]
and we finally obtain the high-temperature one-loop vortex mass
shift formula:

{\normalsize \[
\begin{tabular}{|c|}
\hline \\
$\Delta M_V= -{\hbar m \over 2} \left[
\frac{1}{8\pi\sqrt{\pi}}\cdot \sum_{n=2}^{N_0}\, [\, \sum_{A=1}^4
\, [c_n]_{AA}(K)-c_n(K^{\rm G})]
\cdot\gamma[n-\frac{3}{2},1]+\frac{2 l}{\sqrt{\pi}}
\right]\quad$,$\qquad\beta=\frac{\hbar m}{k_B T}<1$
\\ \\
\hline
\end{tabular}\qquad .
\]}
Numerical integration of the Seeley densities allows us to compute
the heat kernel coefficients. We thus find, by setting $N_0=6$,
the following numerical results for one-loop mass shifts of
superimposed vortices with low magnetic fluxes:
\begin{eqnarray*}
M_V^{l=1}&=&m\left(\frac{\pi v}{e}-1.09427
\hbar\right)+o(\hbar^2)\hspace{2cm} , \hspace{2cm} M_V^{l=2}=2
m\left(\frac{\pi v}{e}-1.08106
\hbar\right)+o(\hbar^2)\\
M_V^{l=3}&=&3 m\left(\frac{\pi v}{e}-1.06230
\hbar\right)+o(\hbar^2) \hspace{2cm} , \hspace{2cm} M_V^{l=4}= 4
m\left(\frac{\pi v}{e}-1.04651\hbar\right)+o(\hbar^2).
\end{eqnarray*}
\section{Summary and outlook}
We have offered a parallel exposition of the derivation of
formulae giving the semi-classical masses of kinks and self-dual
vortices starting from canonical quantization and proceeding
through heat kernel expansions/generalized zeta functions methods.
The treatment of this problem for these respectively
one-dimensional and two-dimensional solitons has thus been
unified. It seems compelling to apply this method to compute the
semi-classical mass of one or other form of Chern-Simons-Higgs
vortices \cite{Dun}.

\end{document}